\documentstyle[12pt]{article}
\begin{document}

\baselineskip24pt
\vbox to 2cm{}

\begin{center}
{\large \bf Molecular dynamics approach:}

{\large \bf from chaotic to statistical properties of compound nuclei}
\vskip1cm

{\bf S. Dro\.zd\.z, J. Oko{\l}owicz, T. Srokowski$^\ast$ and A. Budzanowski}
\vskip1cm

{\em Institute of Nuclear Physics}

{\em ul. Radzikowskiego 152, PL -- 31-342 Krak\'ow, Poland}
\vskip 2cm

{\large ABSTRACT}

\end{center}

Statistical aspects of the dynamics of chaotic scattering in the classical
model of $\alpha$-cluster nuclei are studied.
It is found that the dynamics governed by
hyperbolic instabilities which results in an
exponential decay of the survival probability evolves to a limiting energy
distribution whose density develops the Boltzmann form.
The angular distribution of the corresponding
decay products shows symmetry with respect to $\pi/2$ angle.
Time estimated for the compound nucleus
formation ranges within the order of $10^{-21}$s.
\vfill

\noindent $^\ast$ Present address: GANIL, BP~5027, 14021 Caen Cedex, France
\newpage

In the description of compound nuclei
molecular dynamical approaches [1] (MDA)
generating chaotic behavior appear to provide
an interesting alternative to quantum stochastic methods [2]
based on the random matrix theory.
Indeed, the resulting exponential decay of the classical
survival probability reflects the presence
of Ericson fluctuations [3] as can be seen [4] from the semiclassical energy
autocorrelation function of an S-matrix element. The corresponding unitarity
deficit [5] allows to determine [1]
the probability for the compound nucleus formation.
An important related issue which, however, finds no quantitative documentation
in the literature so far is the problem of statistical properties of
the objects to be interpreted as
compound nuclei formed within the molecular dynamics frame. These properties
are responsible for the decay characteristics such as energetic and angular
distributions of the outgoing particles. The Boltzmann form of the energy
distribution and the symmetry with respect to $\pi/2$ of the angular
distribution are considered to constitute the most convincing signatures
that memory is lost and a certain kind of equilibrium is reached [6].

Because of an explicitly dynamical character MDA offers a very attractive
frame for addressing the related questions.
In particular, does an ensemble of two colliding objects, each composed
of a certain limited number of interacting constituents evolve to some
limiting energy distribution? And, if so, under what conditions this happens,
what is the distribution and what are the time scales involved?
This question differs from the one asked in statistical mechanics where one
assumes that at equilibrium the system will have the most probable
distribution which results in the Boltzmann distribution, the density of which
is $\exp(-E/T)$.
For an isolated system of $n$ interacting particles obeying a nonlinear energy
distribution law, Ulam conjectured [7],
that no matter what the initial distribution
of energy is we have convergence to the exponential distribution. This
conjecture, based on the computer experiments,
was proved later on mathematically [8].
The case of interest for the present considerations corresponds, however,
to open phase space phenomena [9]
and the particles escape from the interaction
region after some time depending on the initial conditions. Is there, then,
enough time for the randomization to occur?

The model [10] specified as a classical limit
of the time-dependent cluster theory [11]
offers an interesting and realistic molecular dynamics scheme, and will be used
here for addressing the above questions. A Van-der-Waals -- type form of the
effective interaction between the elementary constituents, the pointlike alpha
particles in the present case, allows to assign a quite general meaning to
the related analysis.

MDA study of the scattering processes is usually based on the concepts
of the transport theory [12] which, for instance,
allows to determine the classical
survival probability $P_{ij}(E,t)$ for a system to remain in the interaction
region with respect to a $j\to i$ transition. This is a quantity which,
according to the semiclassical considerations determines,
via the Fourier transform,
the energy autocorrelation function $C_{ij}$ of an S-matrix
element, $C_{ij}(\epsilon)=\langle S^*_{ij}(E) S_{ij}(E+\epsilon)\rangle_E$,
and thus makes a link between the quantum and the classical picture.
Chaotic scattering connected with the existence of only unstable periodic
trajectories (hyperbolic chaotic scattering) results in an exponential decay:
$P(E,t) \sim \exp(-\gamma t)$. The resulting autocorrelation function
has a Lorenzian form: $C(\epsilon) \sim \hbar/(\epsilon + i\hbar\gamma)$,
a characteristic of Ericson fluctuations observed in the decay of compound
nuclei. Whether this automatically guarantees appearance of the other
characteristics of the compound nucleus is, however, not immediately obvious.
Actually, the literature presents schematic two dimensional studies
of chaotic scattering on deformed nuclei [13] or on various
three center potentials [14--16] which lead to the
exponential decay of $P(E,t)$ but the angular distribution of outgoing
trajectories is not symmetric with respect to $\pi/2$ angle.

Realistic MDA description of nuclei involves, however,
many more degrees of freedom. In the model considered here both, the target
nucleus and the projectile are composed of the interacting alpha particles
each of them moving in the six-dimensional phase space. Furthermore,
in order to simulate reality and for consistency with the transport theory
the nuclei are defined as the statistical ensembles of elementary constituents.
Each configuration in such an ensemble is constructed so as to ensure
the proper binding energy of a nucleus and its internal linear and angular
momenta zero. Consistently, for a nucleus composed of $n$-bodies any observable
represented by a function $f$ is evaluated by integrals
\begin{eqnarray}
& & \int d{\bf r}_1 \dots d{\bf r}_n \int d{\bf p}_1 \dots d{\bf p}_n
  \delta\bigl(\sum_{i=1}^n {\bf r}_i \bigr)
  \delta\bigl(\sum_{i=1}^n {\bf p}_i \bigr)
  \delta\bigl(\sum_{i=1}^n {\bf l}_i \bigr) \nonumber \\
& & \times
  \delta\bigl(E_B-\sum_{i=1}^n{\bf p}_i^2/2m-V({\bf r}_1,\dots,{\bf r}_n)\bigr)
  f({\bf r}_1,\dots,{\bf p}_n),
\end{eqnarray}
where ${\bf r}_i$, ${\bf p}_i$ and ${\bf l}_i$ denote the position, momentum
and angular momentum of $i$-th alpha particle with respect to the center of
mass of the nucleus.

We begin by studying the time evolution of the energy distribution of
particles for $^{12}$C${}+{}^{12}$C head on reaction and concentrate on the
higher energy part because it eventually drives the decay.
Each of the two colliding
three $\alpha$-particle $^{12}$C nuclei is prepared at separation
according to the above prescription. Firstly, $4\cdot10^5$ scattering events
involving various internal configurations of both nuclei have been generated
on the computer. The time evolution is continued in each case till the
resulting compound system does decay. Thus, for longer times the number
of events which determine the energy distribution of individual particles
decreases. As, however, is shown in Fig.~1 this process clearly
establishes a limiting energy distribution $\rho(E)$
comparatively soon, within the
times of the order of $8-10\cdot10^{-22}\,$s (counting of time begins at the
closest approach distance). Before collision,
the energy distribution of particles is the one representing
ground states of the two nuclei at separation and is well localized.
The initial relative motion boost shifts this distribution to positive
energies.  Early stage of the collision converts energy of the relative motion
into an internal one and, therefore, $\rho(E)$ disperses in the
direction of much higher energies. These high energy particles quickly escape
the interaction region and for the remaining events $\rho(E)$ approaches
an exponential form whose slope allows to define a temperature
like parameter $T$. This, however, is not yet a limiting distribution.
By preserving the exponential form $\rho(E)$ decreases the slope which
reaches the limiting value corresponding to $T=1.3\,$MeV for times of the
order of $10^{-21}$s as can be seen from the time evolution
of the parameter $T$ shown in the lower part of Fig.~1.

The observation of primary interest for our present
discussion is that this time is strongly correlated with time the exponential
decay of the survival probability sets in. This can be concluded
from the upper part of Fig.~2 which shows the number of events $N(t)$ such
that all the particles still remain in the interaction region up to time $t$.
Another important and consistent result,
extracted from these calculations and illustrated
in the lower part of Fig.~2 is that the distribution of the kinetic
energy of the escaping $\alpha$ particles collected from
all the events entering the exponential region $(t>8\cdot 10^{-22}$s)
is also exponential in energy. The slope parameter $T$ describing this
distribution equals $1.5\,$MeV and is thus larger than the one corresponding
to the energy distribution inside the compound system as listed above.
This seems to reflect the fact that the escape
of more energetic particles from the compound nucleus is more probable
due to the Coulomb barrier.
Thus, the `temperature' read over from the escaping particles exceeds
the one inside the nucleus. The fast particles, escaping at an early stage
of the collision (times up to $8\cdot10^{-22}$s) are Gaussian distributed
similarly as they are inside the compound system (Fig.~1).
Finally, we wish to mention, without explicitly demonstrating here the result
that for peripheral collisions of $^{12}$C${}+{}^{12}$C in the decay
channel to the same final configuration we identify the power-law decay
of the survival probability which is typically connected with the existence
of more solid structures (KAM surfaces) in the underlying phase-space [16].
No universal limiting energy distribution exists in this case.

Our study of the statistical properties of a compound nucleus was based so far
essentially on investigation of the temporal aspects of chaotic motion.
More severe test may come from the analysis of spatio-temporal aspects
such as the angular distribution of the decay products. This particular
characteristics is especially interesting in nuclear physics
but the need for studying the spatio-temporal aspects of chaotic motion
is identified [17] also from the more general perspective.
In the present context we are mostly concerned with the conditions under which
the compound
system, in a sense, forgets the way it was formed and as a consequence decays
symmetrically with respect to $\pi/2$ angle in the center of mass frame.
In order to make such a study meaningful one needs to break the mass symmetry
in the entrance channel. For that reason we still consider the same compound
system as before (six $\alpha$-particles) but, this time, produced in the
$\alpha +{}^{20}$Ne reaction. At 20 MeV for this initial configuration
the probability for the compound nucleus formation is much smaller than for
$^{12}$C${}+{}^{12}$C, therefore,
we lower the energy to 15 MeV. The relevant results for two different angular
momenta, $l=0$ and $l=5$ are shown in Fig.~3.
Because of lower energy the dynamics is somewhat slower
and, consequently, the initial stage of the reaction, before the exponential
decay, takes about $5\cdot10^{-22}$s longer than previously.
Still, one observes an impressive coincidence between the behavior of the
survival probability and the form of
angular distribution of the emitted $\alpha$ particles.
Events surviving not longer than $11\cdot10^{-22}$s remember
the initial direction of motion and the emission of the $\alpha$ particles
occurs in the forward direction with much higher probability.
Larger fraction of such a type of events governs the dynamics for $l=5$
than for the central collisions $(l=0)$,
simply because the corresponding transmission
coefficients [1] are smaller for more peripheral collisions.
The angular distribution of all the events decaying after
$t=20\cdot10^{-22}$s shows symmetry with respect to $\pi/2$ angle for
all angular momenta. The dip in the region of $\pi/2$ seen for $l=5$
is the known effect connected with the collective rotation
of the compound system.
The angular distribution of cases decaying for the intermediate times
between $11\cdot10^{-22}$ and $20\cdot10^{-22}$s
also displays the intermediate shapes which reflects the fact that the
transition is gradual. This distribution for $l=5$ is, however, already
closer to the symmetric one than for $l=0$ because of a tendency to regular
orbiting which
reduces the number of particles emitted in forward direction already
at this stage.

The many--body model of nuclear dynamics generating chaotic behavior is
thus able to reproduce not only the probability of the compound nucleus
formation [1] and the correlation width
but also to account for more subtle effects connected with the decay
such as energy and angular distribution of the emitted particles.
The appearance of these characteristics is strongly correlated
with an exponential time dependence of the survival probability.
It is also interesting to notify that all those attributes of compound nuclei
emerge already for the systems with a relatively small number of the
constituents.

We thank Marek Ploszajczak for helpful discussions.
This work was partly supported by KBN Grants No.\ 2~P302~157~04 and
No.\ 2~0334~91~01.

\newpage

{\bf References}
\begin{itemize}

\item[1.]T. Srokowski, J. Oko{\l}owicz, S. Dro\.zd\.z and A. Budzanowski,
Phys. Rev. Lett. {\bf 71}, 2867(1993)

\item[2.]J.J. Varbaarschot, H.A. Weidenm\"uller and M.R. Zirnbauer, Phys.\
Rep. {\bf 129}, 367(1985)

\item[3.]T. Ericson, Phys.\ Rev.\ Lett. {\bf 5}, 430(1960)

\item[4.]R. Bl\"umel and U. Smilansky, Phys.\ Rev.\ Lett. {\bf 60}, 477(1988)

\item[5.]H.L. Harney, F.-M. Dittes and A. M\"uller, Ann.\ Phys. {\bf 220},
159(1992)

\item[6.]J. Benveniste, G. Merkel and A. Mitchell, Phys.\ Rev.\ {\bf C2},
500(1970)

\item[7.]S. Ulam, Adv.\ Appl.\ Math. {\bf 1}, 7(1980),

\item[8.]D. Blackwell and R.D. Mauldin, Lett.\ Math.\ Phys. {\bf 60}, 149(1985)

\item[9.]C.H. Lewenkopf and H.A. Weidenm\"uller, Ann.\ Phys. {\bf 212},
53(1991); Phys.\ Rev.\ Let.\ {\bf 68}, 3511 (1992)

\item[10.]K. M\"ohring, T. Srokowski and D.H.E. Gross, Nucl.\ Phys. {\bf A533},
333(1991)

\item[11.]S. Dro\.zd\.z, J. Oko\l{}owicz and M. Ploszajczak,
Phys.\ Lett.\ {\bf 109B}, 145 (1982);
\item[]W. Bauhoff, E. Caurier, B. Grammaticos and M. Ploszajczak,
Phys.\ Rev.\ {\bf C32}, 1915 (1985)

\item[12.]J.D. Meiss, Rev.\ Mod.\ Phys. {\bf 64}, 795(1992)

\item[13.]A. Rapisarda and M. Baldo, Phys.\ Rev.\ Lett. {\bf 66}, 2581(1991)
\item[]M. Baldo, E.G. Lanza and A. Rapisarda, CHAOS {\bf 3}, No.4 (1993)

\item[14.]C. Jung and H.J. Scholtz, J. Phys.\ A: Math.\ Gen.\ {\bf 20},
3607(1987)

\item[15.]J. Oko{\l}owicz, T. Srokowski and S. Dro\.zd\.z, Nucl.\ Phys.\
{\bf A545}, 479(1992)

\item[16.]S. Dro\.zd\.z, J. Oko{\l}owicz and T. Srokowski, Phys.\ Rev.\ E
{\bf 48}, 4851(1993)

\item[17.]H.~D.~I. Abarbanel, R. Brown, J.~J. Sidorowich and L.~S. Tsimring,
Rev.\ Mod.\ Phys. {\bf 65}, 1331(1993)

\end{itemize}
\newpage

{\bf Figure captions}
\begin{itemize}
\item[1.]{\em Upper part\/}:
Energy distribution of particles for $^{12}$C${}+{}^{12}$C head on reaction at
20 MeV incident energy: before collision (full circles -- solid line is
to guide the eye), at the initial stage ($t=0$ in our time scale) -- when
the relative momentum of the two $^{12}$C becomes zero (squares -- solid line
represents Gaussian fit), at $t=6\cdot10^{-22}$s (triangles -- long dashed
line represents exponential fit), at $t=16\cdot10^{-22}$s (diamonds -- short
dashed exponential fit), and at $t=24\cdot10^{-22}$s (open circles --
dash-dotted exponential fit).\par
{\em Lower part\/}:
Corresponding time dependence of the parameter $T$ describing the slope
of the exponential fits the high energy distribution of particles.

\item[2.]{\em Upper part\/}:
A number of two body events $N$ living until a given time and
leading to an $\alpha $-particle emission
after $^{12}$C${}+{}^{12}$C head on collision at 20 MeV incident energy.
Straight line represents exponential fit for times longer than
$7\cdot10^{-22}$s.\par
{\em Lower part\/}:
Corresponding energy spectra of emitted $\alpha $-particles, collected for
shorter ($t\le 8\cdot10^{-22}$s -- circles) and longer
($t>8\cdot10^{-22}$s -- triangles) times of life
of the composite system (cf.\ {\em upper part\/})
in 0.5 MeV.
Solid line represents Gaussian fit to the former
and dashed line -- exponential fit to the latter.

\item[3.]Relative number of events before $\alpha $-particle emission from
the composite system living up to a time $t$ (on the left) and the yield of
the outgoing $\alpha$-particles (on the right) from the $\alpha + {}^{20}$Ne
head on
reaction, $l=0$ ({\em upper part\/}) and $l=5$ ({\em lower part\/})
at 15 MeV incident energy as a function of the scattering
angle. Triangles represent $\alpha$-particles escaped before
$t=11\cdot10^{-22}$s and circles those escaped after $t=20\cdot10^{-22}$s.
Full line in $l=0$ part represents mean value of the yield.  Distributions for
the intermediate times are indicated by squares.  Error bars correspond to the
statistical uncertainty (square root of the number of events).
\end{itemize}

\end{document}